\documentclass[
aip,
amsmath,amssymb,
reprint
]{revtex4-1}

\usepackage{graphicx}% Include figure files
\usepackage{dcolumn}% Align table columns on decimal point
\usepackage{bm}% bold math

\usepackage[utf8]{inputenc}
\usepackage[T1]{fontenc}
\usepackage{mathptmx}
\usepackage{etoolbox}

\usepackage{color}
\usepackage{graphicx}
\usepackage{amsmath}
\usepackage{amssymb}

\newcommand{\vect}[1]{\mathbf{#1}}

\newcommand{\Vext}{V_\mathrm{fb}}

\usepackage{placeins}
\usepackage{dsfont}
\newcommand{\dif}{\mathrm{d}}
\newcommand{\rring}{r_{\mathrm{ring}}}

\usepackage{placeins}

\usepackage{lipsum}

\makeatletter
\def\@email#1#2{%
	\endgroup
	\patchcmd{\titleblock@produce}
	{\frontmatter@RRAPformat}
	{\frontmatter@RRAPformat{\produce@RRAP{*#1\href{mailto:#2}{#2}}}\frontmatter@RRAPformat}
	{}{}
}%
\makeatother
\begin{document}
	
	\preprint{AIP/123-QED}

\title{Engineering living worms and active crystals with colloidal ``feedback-pullers"}

\author{Sonja Tarama}
\email{starama@fc.ritsumei.ac.jp}
\affiliation{Laboratory of Biological Computation, Department of Bioinformatics, College of Life Sciences, Ritsumeikan University, 1 Chome-1-1 Nojihigashi, Kusatsu, Shiga 525-0058, Japan.}
 
\date{\today}

\begin{abstract}
Using computer simulations, we study the dynamics of colloidal particles with time-delayed feedback interactions. In particular, here we consider ``feedback-pullers", i.e.~colloidal particles that are pulled away from their current position towards an attractive ring centred around their past position. For a single particle, small rings lead to reduced diffusive motion while large rings render activity to the particle. For multiple particles, the particles not only feel their own attractive ring but are also attracted by the rings around all other particles. As expected, for ring sizes larger than the particle diameter, the feedback leads to crystallites whose lattice constant is set by the feedback ring radius. However, here we demonstrate that for long delays (compared to the Brownian time) the colloidal particles start to oscillate around their lattice positions, with the crystallites ultimately collapsing to a close-packed lattice whose lattice constant corresponds to the particle diameter. This effect is caused by the time delay between the particle misplacement within the lattice and the corresponding change in the feedback force. Further, we show that apart from the expected hexagonal crystallites, the time delay may result in the formation of uncommon new states in the case that the ring size is chosen slightly smaller than the particle diameter. Here, particles self-assemble into and move collaboratively as ``living worms'' or as active square-lattice crystallites.
\end{abstract}

\maketitle

\section{Introduction}

The effect of time delay is important in both technology and nature. For technological applications sensing delays may have serious consequences for the system behaviour. E.g.,~in a swarm of robots the sensing delay may change the positional distribution as well as the collective behaviour from segregation to aggregation and clustering \cite{Mijalkov2016, Leyman2018}. Likewise, in nature, the sensory delay in animals \cite{Elzinga2012,More2018}, e.g.~how quick they react to a perceived threat, may indeed be a question of life-or-death.
Previous work addressed the effect of time-delayed perception on the collective motion of animal swarms \cite{Nagy2010,Jiang2017}, finding that delays may enhance or prevent the emergence of clusters and swarms \cite{Piwowarczyk2019,Geiss2022,Pakpour2024,Chen2024} and facilitate coherence \cite{Sun2014}.

In recent years, the effect of feedback on colloidal systems has seen rising interest for both active \cite{Baeuerle2018,Fernandez_Rodriguez2020,Lavergne2019,Khadka2018,Sprenger2022,Chen2023,Wang2023} and passive \cite{Lichtner2010, Loos2014,Loos2017,Loos2018,Kopp2023PRE,Kopp2023EPL,Gernert2015} Brownian particles.
The feedback may lead to propulsion in otherwise passive systems \cite{Gernert2015,Kopp2023PRE,Kopp2023EPL,Bell_Davies2023,Saha2024} and self-organization into larger-scale structures \cite{Juarez2012,Saha2024}. Feedback was also employed as a control mechanism to construct specific interaction rules \cite{Baeuerle2018,Fernandez_Rodriguez2020} or to achieve specific tasks \cite{Yang2020,Li2022,Geiss2019}. 
In these systems, the dependence on a perceived cue can be understood as a self-reinforced confinement \cite{Araujo2023}.

The most obvious way of obtaining active motion from a \textit{delayed} feedback force is by introducing a repulsion from the particle's previous position. The particle is continuously pushed away from its previous position resulting in persistent motion \cite{Kopp2023PRE,Kopp2023EPL}. For many of these ``feedback-pushers'', this can lead to alignment of the particle velocities \cite{Kopp2023EPL} or to self-organization into travelling bands \cite{Tarama2019}. A slightly more complex way to obtain self-propulsion through delayed feedback was recently realized experimentally \cite{Saha2024}. Here, the delayed feedback potential takes the form of an attractive ring placed around the previous position of the particle. Due to the involved time-delay, the particle is situated off-centre in this potential ring, resulting in directed motion towards the potential minimum and, for large attractive rings (of radius larger than the particle diameter), in the formation of particle crystals that did not move. Here, we term such colloids ``feedback-pullers''. 

In this paper, we further explore the collective dynamics of ``feedback-pullers'' and their self-propelling behaviour. 
Self-propulsion has been studied extensively in systems of active Brownian particles \cite{Stenhammar2014,Babel2014} and microswimmers \cite{Stenhammar2017,Qian2017,Oyama2017,Theers2018,Singh2020,Zantop2022}. For the latter, previous studies have found that the collective behaviour depends crucially on the details of the propulsion mechanism, i.e.~differs for pushers (that push out the surrounding fluid in their swimming direction) and pullers (that pull in the fluid in this direction)  \cite{Stenhammar2017,Qian2017,Oyama2017,Theers2018,Singh2020,Zantop2022} due to hydrodynamic interactions between the swimmers. For example, the tendency to undergo motility-induced phase transition (MIPS) changes with the hydrodynamic characteristics \cite{MatasNavarro2014,Theers2018} and optimal steering for task achievement may vary depending on the propulsion mechanism \cite{Daddi_Moussa_Ider2021,Goh2023}. Our study here is similar in spirit although the physical reason for the pulling mechanism is different: in our case it is the prescribed feedback, while it is a hydrodynamic effect mediated by the solvent for the former.

We have previously discussed the case of ``feedback-pushers''\cite{Tarama2019}, demonstrating the formation of travelling bands at time delays similar to the diffusive time scale.
Further, our previous publication \cite{Saha2024} treated the case of  ``feedback-pullers'' with fast but discrete potential updates illustrating how a feedback potential can be used to program freely-choosable interaction potentials in an experimental set-up. Contrarily, here we consider a continuous time-delayed feedback and investigate the change in system behaviour when varying the delay time and potential shape (width and radius). 
Two main results were obtained: Firstly, we show that when using feedback to program crystal phases (that are not densely packed) as we have done in our previous publication, a sharp potential and small time delay are essential.
Secondly, longer delay times (similar to the Brownian time of the particles) lead to the appearance of novel collectively-moving phases. When the potential ring radius is set at values between the particle radius and the particle diameter, the introduced time-delayed interactions lead to moving particle chains (previously reported in \cite{Saha2024}), similar to active worms \cite{Winkler2020,Heeremans2022,Deblais2023}, and a moving square lattice.

The paper is organized as follows: In Section \ref{sec_model} we introduce the underlying equation of motion and simulation methods. We present our results in Section \ref{sec_results} for a single particle (\ref{sec_sinlgep_dynamics}) and for their collective behviour (\ref{sec_collective_dynamics}).

\section{Model and computer simulations}\label{sec_model}
We model a two-dimensional system of $N$ colloidal particles with positions $\vect{r}_i(t)$ ($i=1,...,N$) that are driven by a time-delayed feedback potential force. We describe the particle motion by the over-damped time-delayed Langevin equation
\begin{align}
\gamma\frac{\dif\vect{r}_i}{\dif t}=\vect{f}_i(t)
&+\sum_{j=1}^{N}\vect{F}_\mathrm{fb}\left(\vect{r}_i(t)-\vect{r}_{j}(t-\tau)\right)\nonumber\\
&+\sum_{\substack{j=1\\ j\ne i}}^{N}\vect{F}_\mathrm{WCA}\left(\vect{r}_i(t)-\vect{r}_j(t)\right)\,.\label{eq_eom_manyp}
\end{align}
The left-hand side of eq.~(\ref{eq_eom_manyp}) contains the Stokes drag force with
$\gamma$ denoting the friction coefficient. On the right-hand side the stochastic force $\vect{f}_i(t)$ describes the Brownian thermal motion of the particles.
This random force is Gaussian-distributed with its first 
two moments given by $\langle\vect{f}_i(t)\rangle=0$ and $\langle \vect{f}_i(t)\otimes\vect{f}_j(t')\rangle=2D\gamma^2\mathds{1}\delta\left(t-t'\right)\delta_{ij}$, where 
$D$ is the short-time diffusion coefficient of the particles, $\delta\left(t\right)$ is the Dirac delta function and $\delta_{ij}$ denotes the Kronecker delta.
The feedback forces $\vect{F}_\mathrm{fb}\left(\vect{r}_i(t)-\vect{r}_{j}(t-\tau)\right)$ depend on the distance of the actual particle positions $\vect{r}_i(t)$ to the \textit{previous} positions $\vect{r}_j(t-\tau)$ of \textit{all} particles $j$ (including the self term) where $\tau$ denotes the delay time of the feedback.
We derive $\vect{F}_\mathrm{fb}(\vect{r})$ from a potential $\Vext(r)$ 
as \mbox{$\vect{F}_\mathrm{fb}(\vect{r})=-\nabla \Vext\left(r\right)$} given by a Gaussian ring form 
\begin{align}
\Vext(r)=A\ \exp{\left(-\frac{(r-\rring)^2}{2b^2}\right)}\,,\label{eq_Vext}
\end{align}
of radius $\rring$, potential width $b$ and amplitude $A$. Here, we focus on an attractive ring potential, i.e.~$A<0$. The potential rings are constructed around the former positions of all particles, and each particle experiences force contributions due to all of these rings including their own.

Finally, the equations of motion include direct particle-particle interaction forces
\begin{equation}
\vect{F}_\mathrm{WCA}(\vect{r})=-\nabla V_\text{WCA}(r)
\end{equation}
via a smooth Weeks-Chandler-Anderson (WCA) pair-potential \cite{WCA1971} which takes the form
\begin{align}
V_\text{WCA}(r)=\begin{cases}
	4\varepsilon\left[\left(\frac{d}{r}\right)^{12}-\left(\frac{d}{r}\right)^6\right]+\epsilon\,, &
	r\le 2^{\frac{1}{6}}d\,,\\
	0 &\text{else}\,,
\end{cases}
\end{align}
with the particle diameter $d$ and the potential amplitude $\varepsilon$.

We perform Brownian dynamics simulations in a square simulation box of length $L$ and periodic boundary conditions. The equation of motion, eq.~(\ref{eq_eom_manyp}), is integrated using an explicit Euler scheme with a finite
time step of $\Delta t=10^{-5}\tau_0$ for the particle motion and $\Delta t=10^{-4}\tau_0$ for the feedback update, where $\tau_0=R^2/D$ denotes the Brownian time scale which we define as the time a particle needs to diffuse its own radius $R=d/2$. (For the single particle case, the delay time $\tau$ is used as time scale instead.)
The remaining system parameter values are given in the respective figure captions.

Our simulation protocol is as follows: We start from a random placement of particles. First, the system is briefly equilibrated without any feedback potential for a time $t_\textrm{pre}=10\tau_0$. Next, the positions are recorded during a delay time $\tau$, after which the feedback potential is introduced and the system is relaxed to a steady state (for a duration $t_\textrm{equi}$), now with the feedback. Finally, the particle positions are followed for a long time $t_\mathrm{simu}$ to obtain the results presented in the next section.

\section{Simulation Results}\label{sec_results}
We first consider the dynamics of a single feedback-puller (\ref{sec_sinlgep_dynamics}) before moving on to their collective dynamics (\ref{sec_collective_dynamics}).

\subsection{Single Particle in a Gaussian ring potential}\label{sec_sinlgep_dynamics}
\begin{figure*}[bht]
	\centering
	\includegraphics[width=0.99\textwidth]{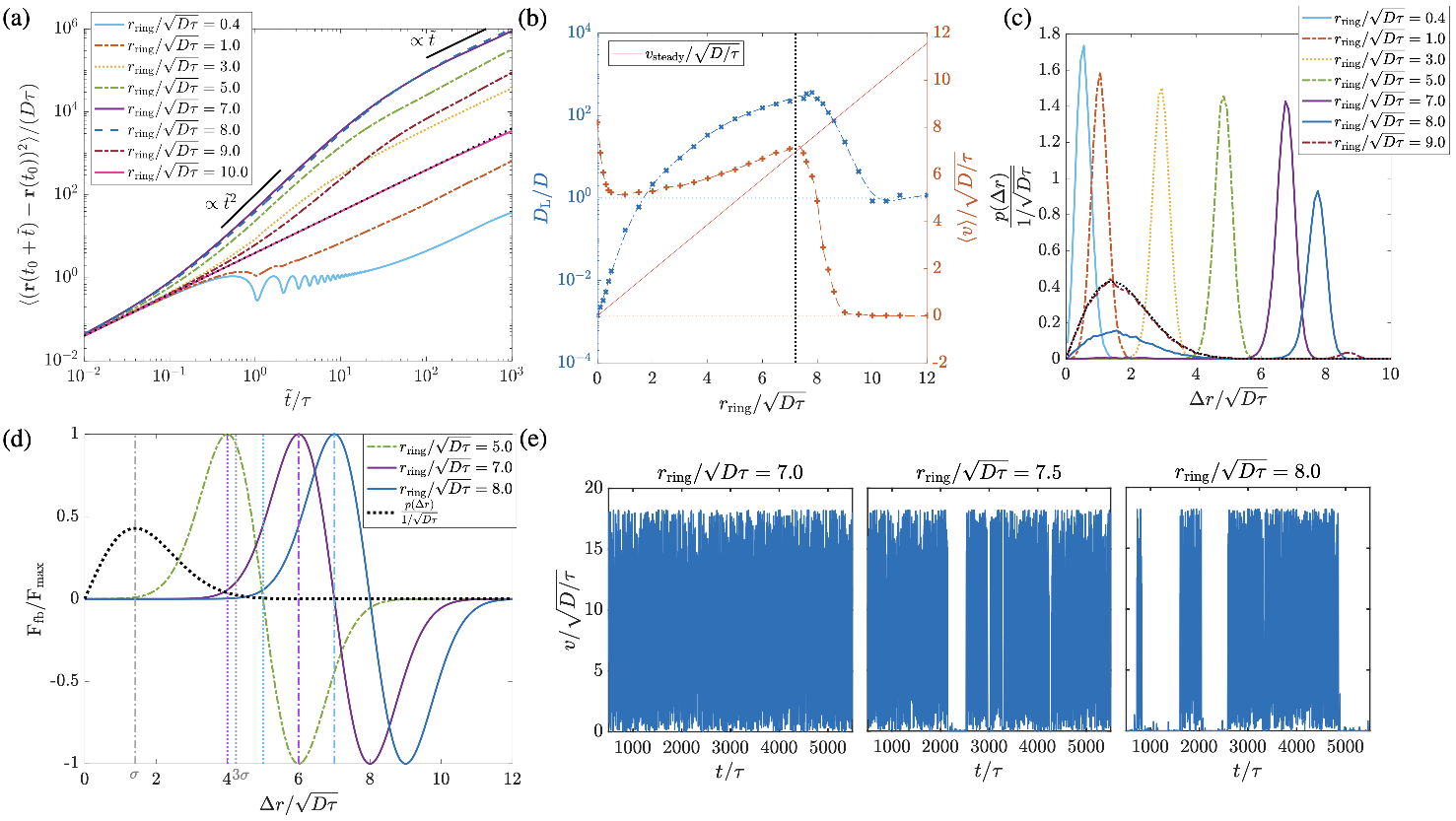}
	\caption{(a) MSD for different ring radii. For small rings the particle is drawn back to its past position, leading to oscillations and a decreased long-time diffusion coefficient $D_\mathrm{L}$ compared to free diffusion (indicated by black dotted line). Increasing the ring radius, the particle is drawn towards the surrounding ring, leading to self-propulsion with a $\propto t^2$ behaviour at intermediate times and an increased $D_\mathrm{L}$. At very large radii, the particle does not feel the potential, leading to free diffusion. (b) Long-time diffusion coefficient and average particle speed for a single particle. Changing the potential radius leads to an increase in the long-time diffusion coefficient $D_\mathrm{L}$ by two orders of magnitude while the average speed changes by a factor of about $1.4$. The vertical black dotted line shows the onset of intermittent propulsion at $\rring/\sqrt{D\tau}\approx7.2$. The red line shows the numerical solution for the speed of a particle moving at a constant velocity $v_\mathrm{steady}$. This steady motion is limited by the strength of the feedback propulsion force. The maximum distance that can be travelled is given by $\Delta r=\mathrm{F}_\mathrm{max}\tau/\gamma$, where $F_\mathrm{max}=-A/b\,\mathrm{e}^{-1/2}$ is the maximum propulsion force from the feedback potential, obtained when the particle sits at $\Delta r=\rring-b$ in the feedback potential. Thus,  $\rring/\sqrt{D\tau}=(\mathrm{F}_\mathrm{max}\tau/\gamma+b)/\sqrt{D\tau}=30\,\mathrm{e}^{-1/2}+1\approx 19.2$ is the maximum ring size for which steady motion in the ring is observed (for the parameters used here, not shown in plot). (c) Probability distribution of the distance between actual and past particle positions $p(\Delta r)$. The dotted black line indicates the Rayleigh distribution in the case of pure diffusion. (d) Rayleigh distribution for a diffusing particle (same as in (c)) and feedback force $F_\mathrm{fb}$ on particle for different ring radii. The overlap between a significant probability of the particle distribution and sufficient feedback propulsion decreases for larger ring sizes. Vertical lines indicate the maximum of the distributions (dash-dotted) and the positions two distribution widths away from these (dotted). (e) Example of particle velocity for different values of the ring radius as a function of time (only part of the simulation is shown). For larger ring values the particle spends exceeding amounts of time in a non-propelling state. [Simulation parameters: $A=-30\gamma D$, $b^2/(D\tau)=1.0$, $t_\mathrm{equi}=500$, $t_\mathrm{simu}=50000$.]}\label{fig_MSD_bs1p0_1p}	\label{fig_DL_v_bs1p0}
\end{figure*} 
For a single particle, the particle dynamics without the feedback potential is purely diffusive. When the feedback potential is introduced as an attractive ring centred around the past position of the particle, the dynamics changes depending on the ring size and time delay to the current position. 

Fig.~\ref{fig_MSD_bs1p0_1p}(a) shows the results for the mean-square displacement (MSD) $\langle(\mathbf{r}(t_0+\tilde{t})-\mathbf{r}(t_0))^2\rangle$ of a single colloidal particle with trajectory $\vect{r}(t)$ in a Gaussian ring feedback potential for rings of different radii as a function of the lag time $\tilde{t}$. Here, $\langle\ldots\rangle$ denotes the average over $t_0$ (equivalent to the average over different realizations of the thermal noise). 
For very small rings ($\rring/\sqrt{D\tau}<2.0$), within the delay time $\tau$ the particle moves farther (by diffusion) than the ring radius. The potential thus draws the particle back to its past position, leading to oscillations in the MSD with period $\tau$ and a reduced long-time diffusion coefficient. The oscillation period can be understood in the following way: Starting from its position at time $t_0$ the particle moves according to the equation of motion. Even without a feedback force, i.e.~in the case that the particle sits in the potential centre, the particle will move due to thermal motion. The feedback ring follows the particle trajectory with a time delay $\tau$. Thus after one delay time $\tau$ (i.e. at $t_0+\tau$), the feedback potential is centred exactly at the particle position at $t_0$, drawing the particle back towards this past position and leading to a minimum in the MSD. Likewise, after $2\tau$, the potential is centred around the position at $t_0+\tau$, pulling the particle towards this position (close to the original position at time $t_0$). The repetition of this process leads to the observed oscillations at multiples of $\tau$ and the particle diffusing on long time scales (MSD $\propto t$) with a strongly reduced diffusion coefficient.

Increasing the ring radius to larger values ($2.0\le\rring/\sqrt{D\tau}\le 9.0$) means that the particle is still situated inside the ring after the delay time $\tau$. Due to the time delay, however, the particle does not sit at the potential centre (at its previous position) but is shifted away from this point. It thus experiences a force from the feedback potential drawing it towards the ring minimum. As the ring position is continuously updated, this leads to propulsion of the particle (indicated by a $\propto t^2$ regime in the MSD). Additional spatial diffusion of the particle leads to an altered angle between the actual and past positions, changing the direction of feedback forces for following times. Thus, on long time scales diffusive motion is observed with an enhanced long-time diffusion coefficient due to the effective propulsion.

For even larger radii ($\rring/\sqrt{D\tau}\ge 10.0$), the particle diffuses within the ring without ever feeling the potential, leading essentially to pure diffusion on all time scales. 

\begin{figure*}[tbh]
	\centering
	\includegraphics[width=0.95\textwidth]{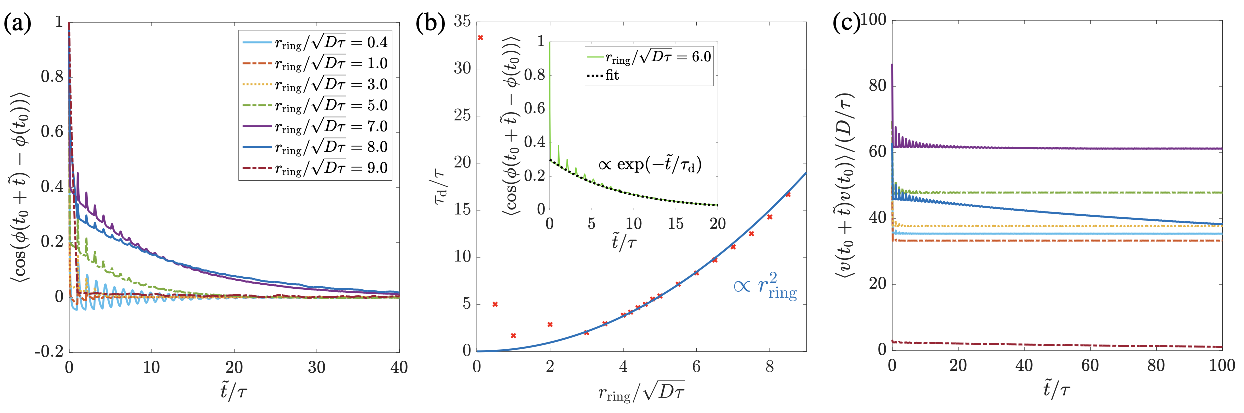}
	\caption{ (a) Correlation of the velocity direction $\phi$. Correlations decay more slowly for larger rings up to a size of about $\rring/\sqrt{D\tau}\approx7.0$. Additional oscillations of period $\tau$ are observed. (b) Decay time of the velocity orientation correlations. The inset shows an example fit for $\rring=6.0$. (The fit ignores the peaked structure.) For intermediate ring sizes, the decay time grows approximately with $\rring^2$. For larger ring radii determining the correlation in orientations becomes difficult due to the intermittent nature of the self-propulsion. (c) Correlation of the velocity amplitude. The particle speed determines the travelled distance within $\tau$ and thus the particle speed at this later point, leading to oscillations of period $\tau$ in the correlation function. Larger velocities are accompanied by longer correlation times of the velocity directions (cf.~(a)). [Simulation parameters: $A=-30\gamma D$, $b^2/(D\tau)=1.0$, $t_\mathrm{equi}=500$, $t_\mathrm{simu}=50000$; see legend in (a) for values of $\rring$ (same for (a) and (c))]}\label{fig_vel_cor_bs1p0}\label{fig_backbone_decay}\label{fig_vmean_1p}
\end{figure*} 

To further investigate the dynamics, we define an effective particle drift velocity due to the systematic feedback force 
\begin{equation}
    \vect{v}=\frac{1}{\gamma}\vect{F}_\mathrm{fb}\left(\vect{r}(t)-\vect{r}(t-\tau)\right)
\end{equation} with its magnitude, the particle speed $v=|\mathbf{v}|$ and direction $\phi$ defined by the angle between the velocity vector and the $x$ axis. Fig.~\ref{fig_DL_v_bs1p0}(b) shows the long-time diffusion coefficient $D_\mathrm{L}$ and the time-averaged particle speed $\langle v \rangle$. The observed propulsion effect at intermediate ring sizes is pronounced. The long-time diffusion coefficient changes by more than two orders of magnitude compared to free diffusion. The particle speed increases towards its maximum at $\rring/\sqrt{D\tau}\approx7$ and falls to zero around $\rring/\sqrt{D\tau}\approx10$ with the diffusion coefficient returning to the free diffusion value $D$.  Fig.~\ref{fig_DL_v_bs1p0}(b) additionally shows the particle speed in a state of steady motion $v_\mathrm{steady}$, fulfilling $\mathrm{F}_\mathrm{fb}(v_\mathrm{steady}\tau)=\gamma v_\mathrm{steady}$, i.e.~the velocity for which the offset between actual and past positions $v_\mathrm{steady}\tau$ leads to a force propelling the particle with the same speed $v_\mathrm{steady}$. For larger rings, the expected constant speed for the case of steady motion $v_\mathrm{steady}$ is larger than the observed average particle speed $\langle v\rangle$. So even though the feedback force could continuously propel the particle at a larger constant speed, the particle slows down beyond $\rring\approx 8.0$. The reason for this behaviour is the particle's diffusive motion. 
Plotting the position of the particle in the feedback potential $\Delta r=|r(t)-r(t-\tau)|$, i.e.~the distance between the actual and past particle position (shown in Fig.~\ref{fig_DL_v_bs1p0}(c)), reveals that this position is not fixed at a value $v_\mathrm{steady}\tau$ but rather, due to diffusion, is broadly distributed within the potential ring.

Moreover, for large rings ($\rring/\sqrt{D\tau}>7.0$) the growth of an additional peak at a smaller distance is observed, which upon further increasing the ring size, grows towards the Rayleigh distribution, i.e.~the distribution expected for a purely diffusing particle after one delay time $\tau$
\begin{equation}
p(\Delta r)=\frac{\Delta r}{\sigma^2}\exp\left(-\frac{(\Delta r)^2}{2\sigma^2}\right)\label{eq_Rayleigh}\,,
\end{equation}
with the distribution width set by $\sigma=\sqrt{2D\tau}$.
This distribution indicates that the particle is in fact not propelled by the ring potential at all.

Fig.~\ref{fig_DL_v_bs1p0}(d) shows the force that the particle experiences as a function of the position within the feedback potential. Additionally, the analytical solution for the probability distribution of the distance for a freely-diffusing particle (eq.~(\ref{eq_Rayleigh})) is plotted. At large ring sizes, a diffusing particle rarely reaches regions of significant feedback force and thus takes a long time to get into or return to a propelled state after leaving the potential ring. The particle thus spends increasingly long times without propulsion. This is also illustrated in Fig.~\ref{fig_DL_v_bs1p0}(e) which shows examples of the instantaneous particle speed for respective simulations of three different ring sizes as a function of time. Here, the propulsion becomes intermittent (similar to run-and-tumble motion \cite{Fier2018,Karani2019,Olsen2024}) between $\rring=7.0$ and $\rring=7.5$ while the particle speed is comparable in all these systems.

\begin{figure*}[]
	\centering
	\includegraphics[width=0.9\textwidth]{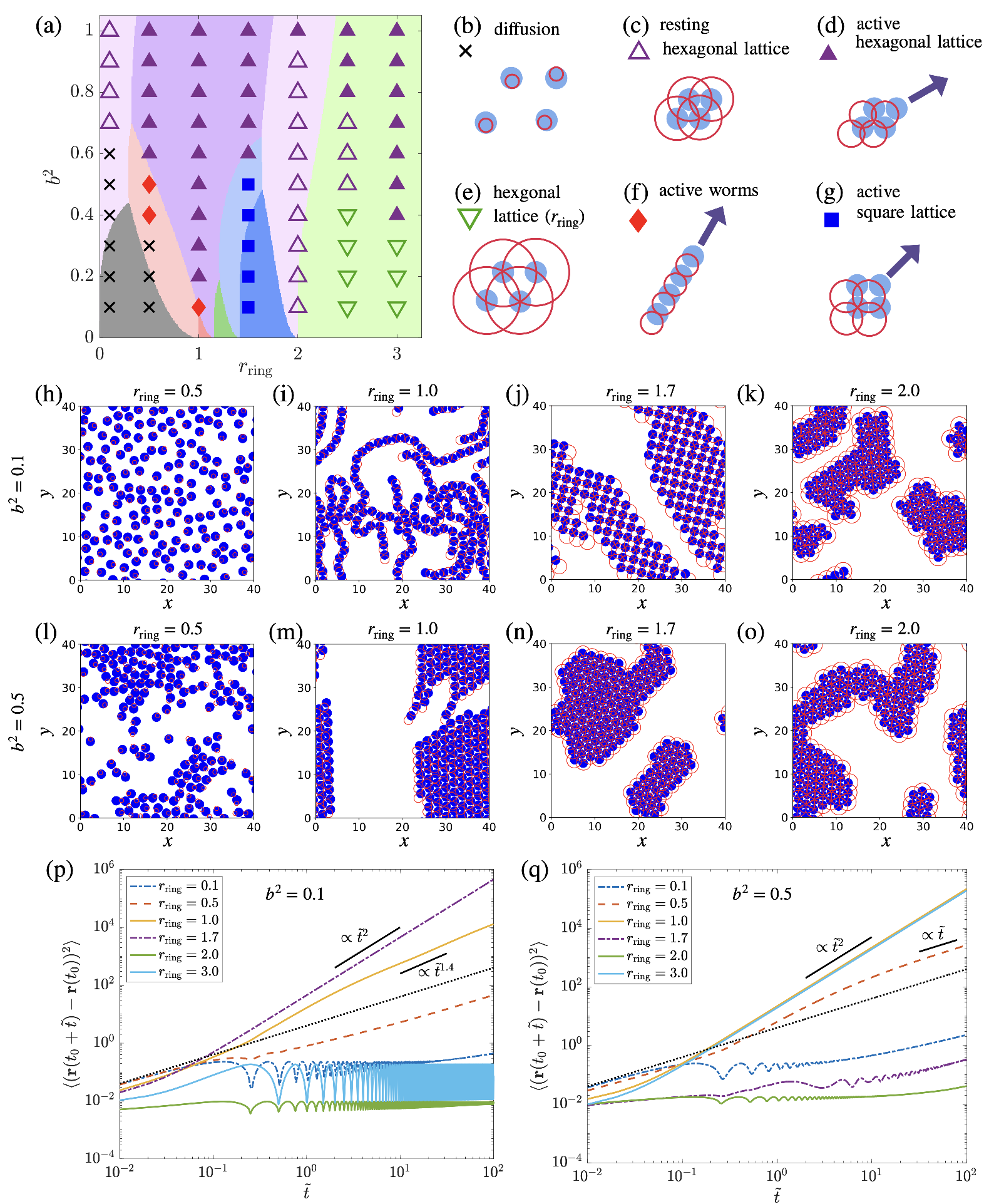}
	\caption{(a) Phase diagram of the system depending on the potential width $b$ and potential ring radius $r_\mathrm{ring}$.  Filled symbols indicate collective motion while empty symbols indicate resting states. The background colour shows a prediction of the system state based on the energies of the states shown in (b)-(g). For red and blue background colour, lighter colour indicates particles at distance $2R$, while darker colour means a structure scaled to fit with $\rring$. For purple and green, darker colour indicates an actively moving lattice. (b)-(g) Sketches of the respective phases. Blue: particles, red: ring-shaped potential minimum of radius $\rring$. (h)-(k) Simulation snapshots for $b^2=0.1$ after $t=550$. The particle dynamics changes drastically from (h) diffusion of individual particles, to (i) chains of moving particles, to (j) a square lattice of moving particles to (k) a resting hexagonal lattice. (l)-(o) Simulation snapshots for a wider potenital $b^2=0.5$ after $t=550$. (p),(q) MSD for the collective dynamics of feedback-driven particles. Chains lead to directed motion but long time sub-ballistic behaviour, the square lattice leads to propelled $\propto t^2$ motion. Close-packed lattices show oscillations of period equal to the delay time $\tau$ ($\rring=2.0$) while lattices at $\rring>2.0$ display oscillations of period $2\tau$ (see MSD for $\rring=3.0$ in (p) as well as Fig.~\ref{fig_crystals}). [Simulation parameters: $t_\mathrm{equi}=100$, $t_\mathrm{simu}=5000$]
	}\label{fig_phase_diagram_snapshots_MSD}
\end{figure*} 

We can broadly estimate the ring size $\rring^*$ at which this happens by considering the overlap between the probability of stay and the region of significant feedback propulsion: The feedback force has its maximum at $\rring-b$ with the distribution width characterized by $b$. The probability distribution for the particle position (eq.~\ref{eq_Rayleigh}) has its maximum at $\sigma=\sqrt{2D\tau}$ and the width characterized by $\sigma$. If we assume the respective contribution to be significant within two widths from the maximum, this leads to the condition $(\rring^*-b)-2b=\sigma+2\sigma$, i.e.~
\begin{equation}
\rring^*=3b+3\sigma\label{eq_rringstar}\,,
\end{equation}
which in our case evaluates to $\rring^*\approx 7.2\sqrt{D\tau}$ and fits quite well with the simulation results (cf.~vertical line in \ref{fig_DL_v_bs1p0}(b) and onset of intermittent behaviour shown in \ref{fig_DL_v_bs1p0}(e)).

Returning to the result for the diffusion coefficient and particle velocity of Fig.~\ref{fig_DL_v_bs1p0}(b), we find that in the same interval ($\rring/\sqrt{D\tau}\approx2.0$ to $\rring/\sqrt{D\tau}\approx 8.0$) that the diffusion coefficient changes by a factor of about $100$, the mean velocity of the particle only changes by a factor of about $1.4$. At first sight this might seem contradictory. However, looking at the orientational correlations $\langle \cos(\phi(t_0+\tau)-\phi(t_0))\rangle$ (shown in Fig.~\ref{fig_vel_cor_bs1p0}(a)), we find that the temporal correlation of the velocity becomes considerably longer when increasing the ring size. In fact, as shown in Fig.~\ref{fig_backbone_decay}(b), we find that the orientational correlations follow an exponential decay with the decay time $\tau_\mathrm{d}$ approximately proportional to the square of the ring size ($\tau_\mathrm{d}\propto r_\mathrm{ring}^2$).
The increase in long-time MSD thus mainly originates from longer correlations in the velocity directions rather than an increase in the particle velocity.

 Fig.~\ref{fig_vel_cor_bs1p0}(c) additionally shows the correlation of the particle speed. As the particle speed determines the distance travelled in the delay time $\tau$ and thus the speed at a later time, oscillations with period $\tau$ are observed. These readily decay, leading to a respective constant long-time limit of $\langle v\rangle^2$ for small and intermediate-sized rings. However, this is not the case for large rings $\rring/\sqrt{D\tau}>8.0$, for which a continuous decline is observed due to the intermittent particle propulsion.

In conclusion, as one would expect, larger rings can pull the particle farther, leading to larger particle speeds and more stable propulsion directions. However, for ring sizes that the particle cannot reach easily by diffusion, propulsion becomes intermittent, reducing its effectiveness. Optimal propulsion is obtained for ring sizes that ensure that the particle still has a sufficient probability to reach the ring by diffusion. An estimate for this ring size is given in eq.~(\ref{eq_rringstar}), showing good agreement with our simulations.

\subsection{Collective dynamics}\label{sec_collective_dynamics}
Next, we examine the effect of the feedback potential on the collective dynamics of ``feedback pullers''.
In the following, lengths are normalized to the particle radius $R$ and times to the Brownian time $\tau_0=R^2/D$. Energies are given in terms of the thermal energy $k_BT\equiv D\gamma$. We drop the units hereafter for ease of notation. We keep the potential strengths $A=-30$ and $\varepsilon=10$ fixed. We use $N=160$ and a particle number density $\rho=0.1$. Unless stated otherwise in the figure caption, we use the delay time $\tau=0.25$.

\subsubsection{Phase diagram}\label{sec_overview}
Fig.~\ref{fig_phase_diagram_snapshots_MSD}(a) shows the phase diagram of the many-particle system for varying potential ring radius $\rring$ and width $b$.
Essentially, apart from diffusion (b), four distinct phases are observed, distinguished by their respective symbols: Hexagonal lattices of lattice constant given by the particle diameter $2R$ (c,d) and potential ring radius $\rring$ (e), respectively, actively-moving worm-like chains of particle (f) as well as an active square lattice state (g). 

Moving through the phase diagram in vertical direction, for sharp potentials (small $b$), increasing the potential radius changes the particle dynamics from reduced diffusion of individual particles, to chains of moving particles, to a square lattice of actively moving particles to a resting triangular lattice.

To understand the origin of the observed particle configurations, we compare the potential energy of a crystal seed of four particles in the respective configurations. The lowest energy state among these is indicated by the background colour in the phase diagram, showing reasonable agreement with the simulation results, thus indicating that the transitions are largely determined by the geometry of the corresponding crystal lattices.
Differences arise due to oversimplifying the observed modes (offsets are not necessarily completely symmetric) as well as neglecting the particle's diffusive motion. In particular, for broader potentials (and not close-packed configurations), particles can still diffusive, destabilizing these phases and thus leading to a tendency to rather form close-packed lattices.

\begin{figure*}[tbh]
	\includegraphics[width=0.95\textwidth]{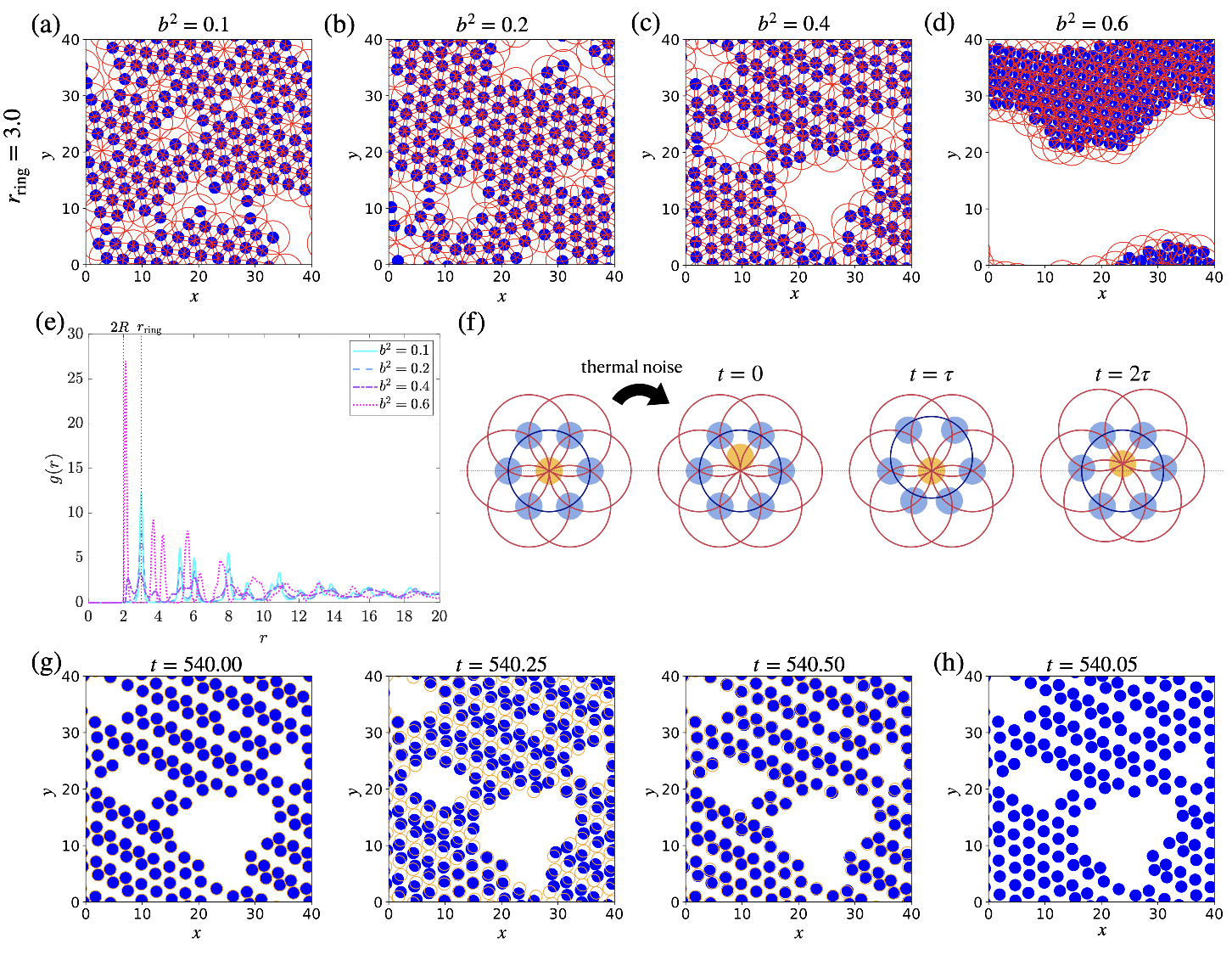}
	\caption{(a)-(d) System snapshots for $r_\mathrm{ring}=3.0$ and different values of $b^2$ after $t=550$. Increasing the potential width $b$, leads to stronger oscillations in the system, with the system ultimately collapsing to a close-packed crystal for large widths. The oscillations are already visible in the MSD of narrow potentials (see Fig.~\ref{fig_phase_diagram_snapshots_MSD}(p)). (e) Pair correlation function $g(r)$ of the crystal lattices. The structure changes from a crystal packed at $\rring$ to a close-packed crystal. Vertical dotted lines indicate the particle diameter $r=2.0$ and the ring radius $r=3.0$, respectively. (f) Schematic sketch of the oscillations in the crystal lattice. Yellow, light blue circles: particles. Red, dark blue rings: minimum of the respective potential rings. An initial displacement (of the yellow particle) due to thermal motion leads to a displacement of the corresponding potential ring (blue ring) a time $\tau$ later. This causes a shift in the positions of the surrounding (blue) particles at this time and therefore a shift in their potential rings a time $2\tau$ after the initial displacement. The shift in the position of the (red) potential rings in turn leads to a displacement of the initial (yellow) particle at this time in the same direction as the initial displacement. In total, the cycle thus leads to oscillations of period $2\tau$. (g) System snapshots showing the oscillations. Orange circles indicate the positions at time $t=540.00$. (h) System snapshot at an intermediate time showing a hexagonal crystal spaced at $\rring$ for part of the system (bottom-left region). Locally, particles oscillate together, but globally oscillations are not synchronized.}\label{fig_crystals}
\end{figure*}

To gain further understanding of the phase diagram, we start by considering the behaviour at small potential radius.
For small rings the particles only feel their own feedback potential which pulls them back to their past position leading to slow diffusive behaviour (Fig.~\ref{fig_phase_diagram_snapshots_MSD}(h)). Increasing the width smears out the potential such that particles feel attraction to their neighbours, leading to clustering into densely packed crystals. 
Increasing the potential radius approximately to the particle radius $R$ leads to the appearance of a new phase: worm-like chains of moving particles (Fig.~\ref{fig_phase_diagram_snapshots_MSD}(i,l)). This state is only long-time stable for very sharp potentials (small $b$) as for broader widths the chains connect at their sides to form a moving cluster (Fig.~\ref{fig_phase_diagram_snapshots_MSD}(m)).
Further increasing the radius (to values $\rring<2$) leads to an active square lattice (Fig.~\ref{fig_phase_diagram_snapshots_MSD}(j)). In this case, the rings are large enough for neighbouring particles to feel the feedback potential. However, the rings are too small for a second particle to sit in the potential minimum around a \textit{non-moving} particle, i.e.~when the feedback potential of a particle is centered around its \textit{actual} position. This results in a crystal state where the particles are constantly in motion and the feedback rings are centered \textit{between} the actual particles. In this case, four particles can fit in a potential ring resulting in a square lattice. Increasing the ring width $b$ leads again to a close-packed hexagonal lattice (Fig.~\ref{fig_phase_diagram_snapshots_MSD}(n)).
\begin{figure*}[bth]
	\includegraphics[width=0.95\textwidth]{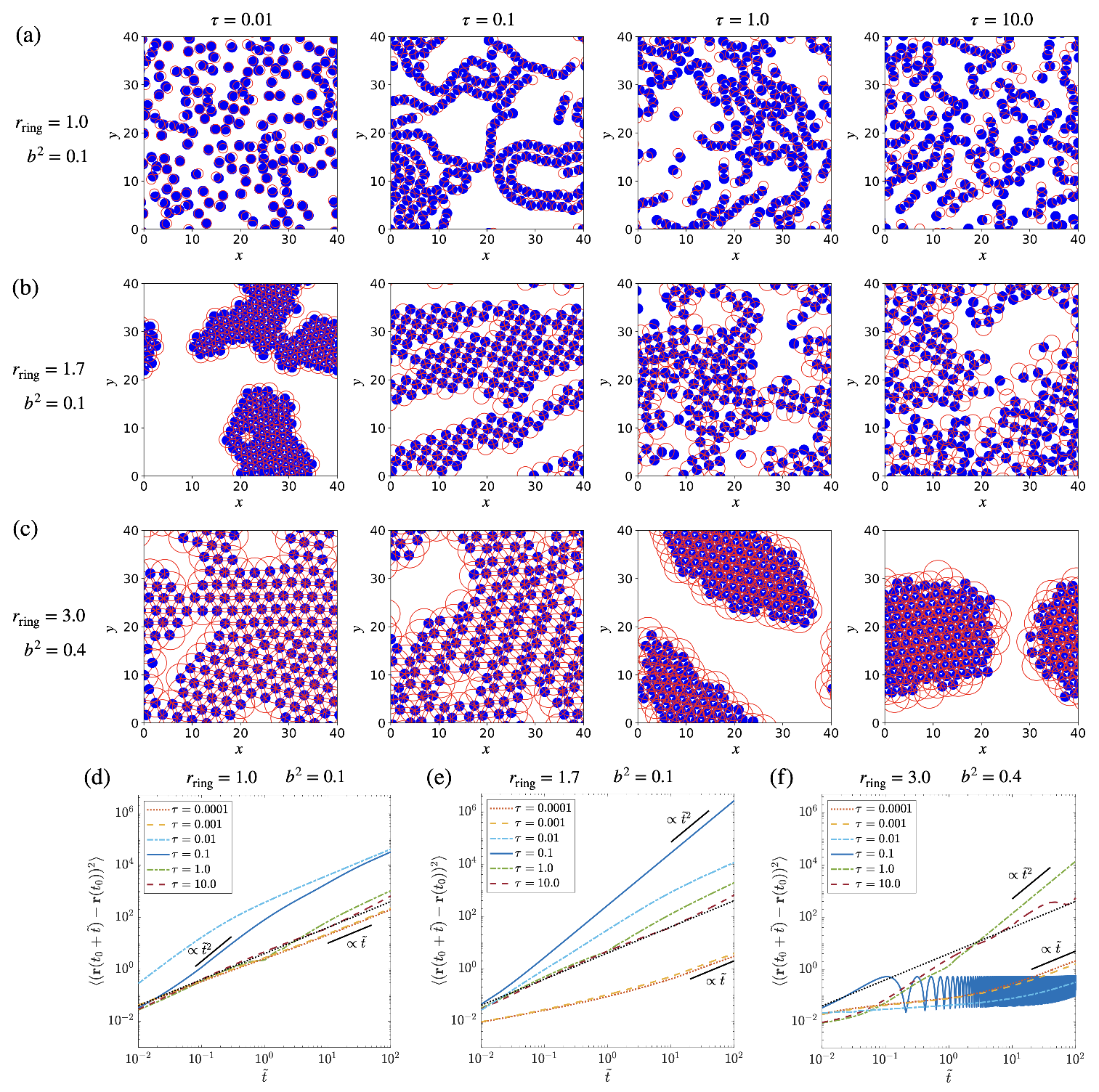}
	\caption{System snapshots for different time delays after $t=2100$. The potential parameters were set to values corresponding to worms (first row), square lattice (second row) and a hexagonal lattice of lattice spacing $\rring$ (third row), respectively. The first two cases rely on a spatial shift between the actual and previous particle positions and thus disappear for short time delay $\tau$. For larger $\tau$, the observed chains become shorter and the square lattice more divided into smaller clusters. Contrarily, the crystal lattice becomes more stable (oscillations disappear), for shorter time delays while at long delays the crystal collapses to a close-packed lattice. (d)-(f) MSD of the three systems for varying delay times $\tau$. The dynamics becomes diffusive both at very short as well as long delays,  while at intermediate delays (a not too small fraction of the Brownian time) propulsion (d,e) and oscillations (f) are observed. Black dotted lines correspond to free diffusion. %data_vel_new_short3_long5/long5_N160_Am30_bs0p4_tau10p0_tch0p001_rring3p0_
	[Simulation parameters: $t_\mathrm{equi}=100$, $t_\mathrm{simu}=5000$; potential update changed to $\Delta t=0.001$ for $\tau=10.0$] }\label{fig_ch_tau}
\end{figure*} 
Finally, for even larger ring radii ($\rring\ge 2$), as one would expect, the particles form a hexagonal lattice (Fig.~\ref{fig_phase_diagram_snapshots_MSD}(k,o)). For $\rring>2$ and relatively sharp potentials (small $b^2$) these will form with the lattice constant equal to the potential ring radius $\rring$. Increasing the potential width however, leads to the collapse of this crystal to a close-packed one that may also move.

To characterize the dynamics of the observed phases, we consider the MSD for the many-particle case in sharp potentials, which is shown in Fig.~\ref{fig_phase_diagram_snapshots_MSD}(p):
At small ring radii ($\rring<1.0$), particles freely diffuse on short time scales, while for longer times the effect of the feedback pulls the particle back to its past position, leading to oscillations of period $\tau$ and at long times to reduced diffusion. Due to the decrease in motion, the particles do not travel far and on the time scales considered, the resulting MSD is identical to the one-particle case. At $\rring=1.0$ the particle-chain state appears with intermediate propelled $\propto t^2$ behaviour followed by a scaling of approximately $\propto t^{1.4}$. The latter scaling is caused by the particle chains wrapping around the system (periodic boundaries), and changes to a $\propto t$-scaling when the system size is increased (not shown).
In the square lattice state at $\rring\approx 1.7$, the particles travel in a cluster leading to a ballistic $\propto t^2$ regime in the MSD.
At larger radii $\rring\ge2.0$ when crystals are formed, the MSD shows oscillations with period equal to the delay time $\tau$ (for $\rring=2.0$) and $2\tau$ (for $\rring>2.0$), respectively. 
Note that for wider potentials (shown in Fig.~\ref{fig_phase_diagram_snapshots_MSD}(q)) the system dynamics changes qualitatively: At small ring sizes subdiffusive behaviour is replaced by superdiffusive behaviour as particles form short chains. Contrarily, for intermediate ring sizes the propulsion is lost ($\rring=1.7$) or newly introduced ($\rring=3.0$) when increasing the ring width, due to the resulting change in particle configuration.

For sharp potentials, a peculiar effect observed in the MSD is the frequency-halving in the oscillations when increasing the ring radius from $2.0$ to $3.0$ (see Fig.~\ref{fig_phase_diagram_snapshots_MSD}(p)). In fact these oscillations are caused by two distinct mechanisms: While the oscillation of period $\tau$ is a direct effect of the delayed confinement and already appears for a single particle, the oscillation of period $2\tau$ is caused by a slightly more intricate mechanism.
Figs.~\ref{fig_crystals}(a-d) show system snapshots for different widths and ring radius fixed to $\rring=3.0$. The oscillations become more prominent when increasing the feedback potential width until the crystal collapses to a close-packed lattice for broad rings. Plotting the pair correlation function (Fig.~\ref{fig_crystals}(e)) reveals that the system is indeed changing from a lattice of spacing $\rring=3.0$ (for $b^2<0.4$) to a spacing $2R=2.0$ ($b^2=0.6$). In between, the system is oscillating between the meta-stable lattice of lattice constant $\rring$ and the close-packed lattice with $b^2=0.4$ showing peaks for both of these distances.
Fig.~\ref{fig_crystals}(f) illustrates the origin of the crystal oscillations. An initial displacement of a particle due to thermal noise leads to a (by $\tau$) delayed displacement of the corresponding potential ring resulting in a shift in the surrounding particle positions. This in turn leads to a spatial shift in the potential around the original position (delayed by $2\tau$), pulling the first particle away from its original position again. Thus, oscillations of period $2\tau$ emerge.
Similar oscillations were previously reported for molecules of Brownian particles bound by a delayed harmonic potential \cite{Geiss2019,Khadka2018}.

Fig.~\ref{fig_crystals}(g) shows the resulting oscillations in our simulations. Here, rows of particles in the crystal lattice oscillate together around their lattice positions. Fig.~\ref{fig_crystals}(h) additionally shows the intermediate equally-spaced lattice in \textit{part of} the system, illustrating that the observed oscillations are only locally but not globally synchronized.

\subsubsection{Dependence on the delay time}

Finally, we investigate how the time delay influences the system dynamics. 
Figs.~\ref{fig_ch_tau}(a)-(c) show simulation snapshots for three characteristic phases, given by worms, a square lattice and a hexagonal lattice when changing the delay time. Figs.~\ref{fig_ch_tau}(d)-(f) give the MSD of the respective systems. The results for the second column ($\tau=0.1$) are qualitatively the same as for the delay used before ($\tau=0.25$) so we use this case as a reference when considering the effect of the time delay.

Firstly, for the particle chains (first row), the number and length of the particle chains is significantly reduced for both small ($\tau<0.01$) and long delays ($\tau > 1.0$) with the MSD (d) showing diffusive behaviour in these cases. Contrarily, at intermediate delays, chains form and the dynamics becomes ballistic on time scales similar to the delay time ($\tau=0.01,0.1,1.0$).

For the active square lattice (second row), decreasing the time delay leads to the formation of a densely-packed hexagonal cluster ($\tau=0.01$). In fact, for short time delays the potential ring is centred approximately at the actual particle centre and can thus be understood as a weak inter-particle attraction. In the limit of long delay, the square lattice is maintained but smaller crystals are formed ($\tau=1.0, 10.0$). The MSD (e) reveals ballistic behaviour at intermediate delay times ($\tau=0.1$) and increased diffusion for both slightly shorter ($\tau=0.01$) and longer delays ($\tau>1.0$) while very small delays ($\tau=0.001,0.0001$) show reduced diffusion due to the formation of the hexagonal lattice.

For the crystal (third row), no oscillations are observed for short time delays ($\tau<0.1$), while for long delays the crystal collapses to a close-packed (active) lattice ($\tau=1.0,10.0$). We thus rationalize that longer delay times destabilize the formed crystal lattice, due to the time delay between a particle becoming misaligned and the corresponding change in the feedback forces.

In conclusion, we find that crystals are stable in the instantaneous interaction limit ($\tau\to0$) while the square lattice and particle chain phases require an offset between the actual and previous particle positions and thus disappear in this limit. Contrarily, long time delays reduce ordering in all of these systems.

\section{Conclusions} \label{sec_conclusion}
We have characterized the collective dynamics of ``feedback-pullers", i.e.~colloidal particles that self-propel by pulling themselves towards an attractive ring centred around their past position. 

For a single particle, this self-propulsion leads to ballistic behaviour on intermediate time scales and an increased diffusion in the long-time limit.
For many particles, our results revealed the existence of uncommon new phases given by an active square lattice crystal state as well as chains of moving particles. Additionally, hexagonal crystal lattices were observed whose lattice constant can be controlled at wish, a possibility which has already been realized experimentally in our previous publication \cite{Saha2024}. For this case, our results presented here show that the time delay needs to be kept small compared to the diffusive time scale $\tau_B$, as otherwise oscillations appear. On the contrary, to obtain the novel active square lattice and worm-like particle-chain states, time delays should be kept at values only slightly smaller than $\tau_B$.

Our results suggest that colloidal feedback control can be used to create new travelling crystals with open structures as well as living polymers. As demonstrated in this work, the lattice structure and spacing can efficiently be tuned by the shape of the feedback potential. This is of utmost importance for the fabrication of controlled active elements needed for next-generation microswimmers and for the design of functionalized materials with novel structural and dynamical properties.

%\FloatBarrier
\begin{acknowledgments}
	We thank Mitsusuke Tarama and Hartmut Löwen for proofreading of and helpful comments on the manuscript. We thank Debasish Saha and Stefan U. Egelhaaf for helpful discussions.
\end{acknowledgments}

\FloatBarrier

%\nocite{*}
\bibliography{references}

\end{document}